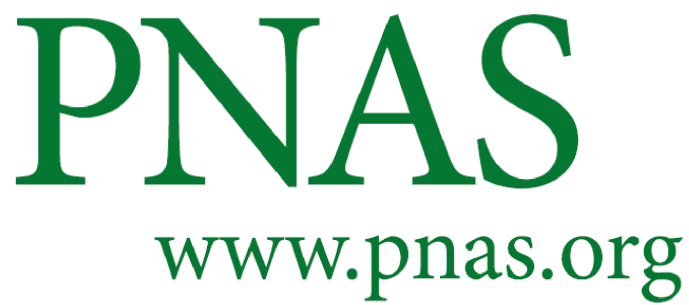

**Main Manuscript for**

Concentration Gradients in Evaporating Binary Droplets Probed by Spatially Resolved Raman and NMR Spectroscopy

Alena K. Bell[1 †], Jonas Kind[3 †], Maximilian Hartmann[2], Benjamin Kresse[4], Mark V. Höfler[4], Benedikt B. Straub[5], Günter K. Auernhammer[5,6], Michael Vogel[4], Christina M. Thiele[3#], Robert W. Stark[1*]

[1] Physics of Surfaces, Institute of Materials Science, Technical University of Darmstadt, Alarich-Weiss-Str. 16, 64287 Darmstadt, Germany

[2] Institute for Nano- and Microfluidics, Technical University of Darmstadt, Alarich-Weiss-Str. 10, 64287 Darmstadt, Germany

[3] Clemens-Schöpf-Institute for Organic Chemistry and Biochemistry, Technical University of Darmstadt, Alarich-Weiss-Str. 16, 64287 Darmstadt, Germany

[4] Institute of Condensed Matter Physics, Technical University of Darmstadt, Hochschulstr. 6, 64289 Darmstadt, Germany

[5] Max Planck Institute for Polymer Research, Ackermannweg 10, 55128 Mainz, Germany

[6] Leibniz-Institut für Polymerforschung, Hohe Straße 6, 01069 Dresden, Germany

† Both authors contributed equally

# corresponding author, cthiele@thielelab.de

* corresponding author, robert.stark@tu-darmstadt.de,

**Email:** cthiele@thielelab.de and robert.stark@tu-darmstadt.de,

**Author Contributions:** C.M.T., G.K.A., M.V. and R.W.S. conceived the idea. A.K.B. designed and conducted the Raman experiments. Data analysis and data visualization of the Raman data were carried out by A.K.B.. J.K., B.K. and M.V.H. designed and conducted the NMR experiments. M.H. designed and conducted the CA experiments and the coating of the substrates. Data analysis and data visualization of the CA data were carried out by M.H.. B.B.S. designed and conducted the APTV experiments. Data analysis and data visualization of the APTV data were carried out by B.B.S.. A.K.B., J.K., C.M.T. and R.W.S. prepared the original draft. All authors discussed the results and revised the manuscript.

**Competing Interest Statement:** The authors declare no competing interests.

**Classification:** Physical Sciences, Applied Physical Sciences

**This PDF file includes:**

    Main Text
    Figures 1 to 4

**Abstract**

Understanding the evaporation process of binary sessile droplets is essential for optimizing various technical processes, such as inkjet printing or heat transfer. Liquid mixtures whose evaporation and wetting properties may differ significantly from those of pure liquids are particularly interesting. Concentration gradients may occur in these binary droplets. The challenge is to measure concentration gradients without affecting the evaporation process. Here, spectroscopic methods with spatial resolution can discriminate between the components of a liquid mixture. We show that confocal Raman microscopy and spatially resolved nuclear magnetic resonance (NMR) spectroscopy can be used as complementary methods to measure concentration gradients in evaporating 1-butanol/1-hexanol droplets on a hydrophobic surface. Deuterating one of the liquids allows analysis of the local composition through the comparison of the intensities of the CH and CD stretching bands in Raman spectra. Spatially resolved NMR spectroscopy is used to measure the composition at different positions of the droplet. Confocal Raman and spatially resolved NMR experiments show the presence of a vertical concentration gradient as the 1-butanol/1-hexanol droplet evaporates.

**Significance Statement**

Imagine you spill your drink and miss some spots when cleaning up. The next morning you notice that the stains look quite different on different surfaces. What has happened? In droplets of liquid mixtures, the components evaporate at different rates which leads to gradients in concentration and surface tension. These gradients can cause, for example, so called Marangoni flows which in turn affect the evaporation process. To better understand evaporation induced liquid flows, the concentration gradients have to be measured without disturbing the liquid. Marker molecules might be surface active or even may affect the evaporation process. We report here on marker-free and contact-less measurements of concentrations by spatially resolved Raman and NMR spectroscopies in evaporating binary droplets.

**Main Text**

**Introduction**

Evaporating droplets occur in various contexts such as inkjet printing (1, 2), heat transfer or daily phenomena such as drying coffee stains. (3, 4) In many applications, such as painting (5), cleaning, gluing, or printing (6), where liquid mixtures are used, the evaporation of a droplet is a rather complex process because the concentration profile within the droplet varies over time. To improve the controllability and predictability of the technical processes, it is essential to characterize the transport phenomena during the drying process. To this end, the measurement of the droplet composition is a crucial element and has to be carried out with sufficient spatial and temporal resolution. In particular, spectroscopic methods are promising tools for contact-less concentration measurements of liquid mixtures.

The evaporation of a droplet is governed by physical properties such as surface tension, density, vapour pressure, and boiling temperature. Additionally, concentration gradients can evolve in liquid mixtures. (7) These gradients are driven by thermal gradients due to the enthalpy of evaporation or heating or by surface tension gradients induced by preferential evaporation of one component. The evaporation rates of the components can vary over the droplet surface. For sessile droplets with contact angles smaller than 90°, for example, the evaporation rates are higher at the three-phase contact line. (8) These thermal or surface tension gradients usually induce flow inside the droplet called Marangoni flow. This flow leads to concentration gradients across the droplet. Due to the nature of the Marangoni flow inside of a spherical sessile droplet, no concentration gradients are expected at the centre axis (perpendicular to the substrate), whereas further outside, such vertical gradients can occur. (9-12) The direction of the gradient depends on the density and surface tension. A direct application of this principle is, for instance, Marangoni cleaning in semiconductor technology. (13)

The investigation of the composition of tiny sessile drops, as they occur in inkjet printing or other technical processes, poses a challenge because the typical length scales of interest are smaller than the capillary wave-length. In bulk samples, the composition can be examined in a straight-forward manner with chromatographic methods such as gas chromatography (GC) and high-performance liquid chromatography (HPLC) or spectroscopic methods such as nuclear magnetic resonance (NMR) spectroscopy, infrared (IR) spectroscopy and Raman spectroscopy. However, for the investigation of sessile droplets, a high spatial and temporal resolution is required. For this purpose, confocal Raman spectroscopy and spatially resolved NMR spectroscopy are powerful tools. For both techniques, concentration determination is straight-forward if at least two signals of the components of interest are baseline separated. NMR is intrinsically calibration free, whereas Raman spectroscopy requires calibration through reference experiments. (14-16) Both approaches allow the quantification of concentration gradients in sessile droplets, as is shown here.

In Raman microscopy, good spatial resolution can be achieved in a confocal setup. The components of mixtures can be distinguished via specific vibrations for different functional groups or through a careful analysis of the Raman signals in the fingerprint region (<1500 cm$^{-1}$). For example, binary mixtures of ethanol and water can be characterized in a straight-forward manner. (15) If, however, both liquids have a similar chemical structure, the discrimination of the components might be hampered by signal overlap in the CH stretching region (2800-3000 cm$^{-1}$); e.g., in such cases, Raman signals in the fingerprint region (<1500 cm$^{-1}$) might be used for the identification of the species. However, these signals often provide a poor signal-to-noise ratio, which makes large integration times necessary. Thus, the image rate or resolution is so low that even slow diffusion processes are hardly resolved. Here, Raman stable isotope probing (SIP), which has been developed to monitor metabolic processes in microbiology, offers a solution. (17) The basic idea of Raman SIP is to replace the proton in the CH with deuterium in one of the mixture components such that the CD stretching region occurs at roughly $1/\sqrt{2}$ times the CH stretching and falls into a region with very weak or even without signals from the protonated liquid component. Thus, the concentration in a binary mixture can be calculated in a straight-forward manner from the ratio of the integrated Raman intensities $I_{CD}/I_{CH}$ of the respective stretching vibrations.

Compared to Raman microscopy, where localization is achieved by scanning the focal point across the region of interest, in NMR experiments, localization is achieved by using magnetic field gradients. Usually, one avoids phase boundaries (especially liquid-gas interfaces) in NMR experiments because they disturb the magnetic field homogeneity and reduce the spectral quality in terms of line shape and baseline separation of the resonances. Nevertheless, it has been shown that magnetic resonance imaging (MRI) can be used to characterize freezing water droplets (18), the infiltration of water into asphalts (19), and the evaporation of sessile droplets from porous surfaces. (20-22) Additionally, NMR can be used to quantify the composition of binary droplets during evaporation. (23)

Thus, the use of both complementary approaches to characterize evaporating binary droplets may be beneficial. In this article, we discuss the capabilities of Raman SIP and NMR techniques to analyse the evolution of the composition of an evaporating sessile binary droplet. As a model system, a binary mixture of 1-butanol and 1-hexanol was used. This mixture shows a low volatility such that the evaporation process can be captured with both Raman and NMR spectroscopies. With Raman spectroscopy, it was possible to observe concentration gradients of 1-butanol-$d_9$ over the height of the droplet during evaporation. NMR techniques were examined in terms of the capability to observe the evaporation of 1-butanol and yield time-dependent droplet composition with spatially resolved $^1$H-NMR spectra. Furthermore, the contours of the evaporating droplets were tracked by optical measurements to characterize the time-dependent changes in the droplet dimensions.

**Results**

**Binary Droplets**

Five different systems were considered in this work: pure 1-butanol, pure 1-butan-$d_9$-ol (referred to as 1-butanol-$d_9$), pure 1-hexanol, and mixtures of 1-butanol/1-hexanol and 1-butanol-$d_9$/1-hexanol. The relevant parameters of the substances are given in Table S1. The evaporation of binary droplets composed of 1-butanol/1-hexanol was investigated. Both alcohols have lower surface tensions than water, so a droplet of a 1-butanol/1-hexanol mixture has a smaller contact angle than a droplet of pure water. In the binary mixture, 1-butanol was expected to be the more volatile component because it has a lower boiling point and higher vapor pressure. As a result, a sessile binary droplet composed of 1-butanol and 1-hexanol with a contact angle below 90° shows an enhanced evaporation rate for 1-butanol at the three-phase contact line. Thus, flows in the droplet are driven by concentration gradients and gradients of the surface tension (Marangoni flow).

For Raman imaging, isotope-labelled fluids were used. The $^2$H isotopologues of both alcohols are commercially available, so lengthy deuteration procedures were not needed, and 1-butanol-$d_9$ was used for Raman SIP. The peak of the CD stretching vibration served as a marker. Due to the approximately doubled mass of the deuteron vs. the proton, the resonance frequency of the CD vibration is decreased by a factor of roughly $\sqrt{2}$ compared to the CH vibrations. (24-31) It is expected that the evaporation behaviours of different isotopologues of an alcohol differ depending on the degree and position of deuteration, as density, hydrogen bonding and intermolecular dispersion interactions are affected. Hence, we used 1-butanol-$d_9$ instead of 1-butanol-$d_{10}$ to retain comparable intermolecular hydrogen bonding interactions. Due to the effect of deuteration (Table S1) on the density and intermolecular dispersion interactions, 1-butanol-$d_9$ has a slightly lower vapor pressure than 1-butanol-$d_{10}$. Hence, 1-butanol-$d_9$ is expected to evaporate somewhat slower than 1-butanol-$d_{10}$. This difference is much smaller than the difference from 1-hexanol. We considered this difference in evaporation rates negligible.

**Confocal Raman Stable Isotope Imaging**

Raman measurements were taken as depth image scans in the *xz*-plane and were repeated continuously to track the evolution of the concentration profile. The measurements were slightly off the droplet centre to monitor the vertical concentration gradient over the drop height. At the drop centre axis, the concentration should be height independent, while the interplay between component-dependent evaporation rates and Marangoni flows sustains a concentration gradient. (9-12)

First, reference measurements were performed with pure substances and mixtures of protonated substances (see SI figures S4-S5). From the reference experiments, the timescales of evaporation could be extracted for the single components and mixtures of substances. The concentration profiles, however, could not be measured with sufficient resolution because the most prominent Raman bands at 2900 cm$^{-1}$ overlapped.

By using 1-butanol-$d_9$ overlapping C-H stretching bands are avoided. The Raman spectra of pure 1-butanol-$d_9$ and 1-hexanol are given in figure S3, as well as the Raman spectrum of the mixture of both substances. As described above the C-H stretching band of 1-Hexanol (2800-3000 cm$^{-1}$) and the C-D stretching band of 1-butanol-$d_9$ (2000-2400 cm$^{-1}$) are baseline separated. Hence, the ratio of the integrated band intensities can be used as a measure for the concentration (27) by using a calibration curve. (14-16)

The measurements show different evaporation behaviours for the two alcohols of different chain lengths. 1-Butanol and 1-butanol-$d_9$ have lower boiling temperatures, higher vapor pressures and higher surface tensions -they evaporated within 30 min. 1-Hexanol has a lower vapor pressure and thus evaporated much more slowly, in approximately 190 min. The mixture of both substances had an intermediate evaporation time of almost 120 min.

Because the scans were taken through the liquid-gas interface, variations in the optical path had to be considered. One might think that refraction must be considered because the droplet acts as an additional lens in the optical path. A sketch to visualize the effect on a vertical line in a scan is given in the supporting information (figure S8). Briefly, according to Snell's law, the transition from air ($n_{air}$=1) to the curved surface of 1-butanol ($n_{but}$=1.3988) or 1-hexanol ($n_{hex}$=1.418) causes refraction of the laser focus in the direction of the drop centre and changes the focal length of the system. This leads to a distortion of the image (the image is then compressed and slightly curved). In the centre of the drop, i.e., close to the optic axis, compression or stretching prevails, while bending distortion can be neglected. For this reason, the measurements were conducted close to the centre of the drop (optical axis) so that the scattering effects could be safely neglected.

The intensity distribution of the peaks corresponding to the substances are given in the depth scans, shown in figure 1(a) and 1(b). The depth scans are raw data, i.e., they have not been corrected for the optical distortion. From left to right, the temporal sequence of scans of the same drop is shown. To calculate the concentration from these intensity distribution images, the calibration curve, given in figure S6(a), is needed. In figure 1(c), the concentration profile in every scan is given, and in figure 1(d), the concentration gradient over the experimental time is given.

From the confocal Raman scans in figures 1(a) and 1(b), the droplet height can be estimated from the interface between liquid and vapor (top of the stripes). Plotting the height of the 1-hexanol droplet over time reveals that the height decreased in a non-linear manner, which is explained by evaporation in stick-slip mode, as discussed in the previous section. Initially, the height of the droplet decreased rapidly, and this process slowed. The initial fast decrease was correlated to the evaporation of the more volatile component (1-butanol or 1-butanol-$d_9$) of the 1-butanol-$d_9$/1-hexanol mixture. After most of the butanol component had evaporated, the 1-hexanol most likely evaporated as the pure substance. Additionally, a non-linear, wave-like decrease in height was observed due to a stick-slip evaporation mode.

This observation is in agreement with data from contact angle goniometry data (see SI). The concentration gradient for the 1-butanol-$d_9$/1-hexanol mixture can be calculated from the intensity distribution images in figures 1(a) and 1(b). The intensity distribution images (figure 1(a) and 1(b)) show a faster 1-butanol-$d_9$ depletion in the upper part of the droplet than near the substrate. This distribution changed after the first four scans (34 min) when the more volatile 1-butanol-$d_9$ had evaporated such that only a 1-hexanol signal was detected. The highest 1-butanol-$d_9$ fraction ($55 \pm 15$ mol%) was measured near the substrate in the first scan. At the liquid/gas interface, the 1-butanol-$d_9$ fraction was $40 \pm 12.5$ mol%. The vertical concentration profile is given in figure 1(c), and the evolution of the concentration gradient is given in the graph in figure 1(d). In the first scans, a gradient of $0.024 \pm 0.001$ mol%/µm was observed, which decreased to nearly zero ($0.003 \pm 0.003$ mol%/µm) after the first 5 scans. This was also observed in the intensity images and the concentration graph. The results also show that the 1-butanol-$d_9$ concentration decreased with time in the whole droplet. The lower detection limit of the 1-butanol-$d_9$ concentration was $12 \pm 14$ mol% due to the low signal-to-noise ratio. Additionally, the inspection of the Raman spectra after the first five or six scans indicated that 1-butanol-$d_9$ evaporated below the detection limit.

These concentration gradients in the bulk give rise to concentration (and probably temperature) gradients along the droplet surface. These surface gradients lead to surface tension gradients and induce Marangoni flows. To verify the Marangoni flows, astigmatism particle tracking velocimetry (APTV) measurements were performed. From these measurements, a flow from the drop centre along the drop surface towards the three-phase contact line and back to the bulk was observed, proving the existence of surface tension gradients. The details of these results are given in supplementary information with the images in figure S15.

**Concentration Determination Using Spatially Resolved Nuclear Magnetic Resonance Spectroscopy**

The evaporation of 1-butanol/1-hexanol droplets was also observed using (non)spatially resolved NMR techniques as an independent and complementary characterization method. In contrast to Raman spectroscopy, it was not necessary to use deuterated 1-butanol. Both alcohols have similar chemical shifts for all resonances in isotropic bulk samples (figure 2). Further complications arose for small droplets on surfaces, as the magnetic field homogeneity is influenced by the large liquid-gas interface, so the resulting line width in NMR spectra did not allow for direct analysis of the composition via integration of the resonances belonging to the two different species. Nevertheless, the molar ratios of 1-butanol and 1-hexanol in bulk samples and droplets could be determined from $^1$H NMR spectra by comparing the integrals of $CH_2$ and $CH_3$ resonances. For pure 1-butanol, a ratio of 4:3 was expected, and for pure 1-hexanol, a ratio of 8:3 was expected (figure 2). Hence, the fraction could be calculated with equation 1.

$$X_{1-hexanol} = 0.75 \cdot \left(\frac{I_{CH_2}}{I_{CH_3}}\right) - 1$$

$$X_{1-butanol} = 1 - X_{1-hexanol}$$

In contrast to bulk samples, for droplets, artefacts due to differences in susceptibility are expected, which should be smaller at lower magnetic fields. Thus, molar fractions of 1-butanol and 1-hexanol in evaporating droplets of a 50:50 mol% mixture were examined with non-localized NMR spectroscopy at a low field. As a result, the time-dependent bulk composition of the droplet during evaporation was obtained (See SI). The concentration of the more volatile component $X_{But}$ decreased during evaporation but did not reach zero. Rather, it appeared to saturate at a concentration of slightly less than 20 mol%. Below that, errors in measurement did not allow further conclusions.

To minimize the effects of the liquid/gas interface on the line shape and width observed, PRESS (Point RESolved Spectra, in the following described as localized spectroscopy) NMR spectra were acquired within the droplet instead of measuring the whole droplet. The voxel size and position were chosen in a way that no interphase was within the region of interest (figure 3).

To investigate whether it is possible to locally detect a concentration gradient during evaporation, three small voxels (0.2 mm x 0.2 mm x 0.2 mm) positioned above each other were selected so that spectra were obtained over the entire drop height of initially approx. 650 µm (see coloured boxes in figure 3). In contrast to non-spatially resolved NMR spectra, in the PRESS spectra acquired with this voxel size, the overall linewidth was significantly smaller. As a result, the $CH_2$ and $CH_3$ resonances were baseline separated (see figures S8 and S9). Hence, the molar fractions of 1-butanol and 1-hexanol could be calculated using equation 1.

A faster decrease in butanol concentration was observed for the upper voxel than for the middle and lower voxels (figure 4). For the topmost voxel, the 1-butanol fraction decayed to zero within 500 min. In the bottom voxel, 1-butanol could be detected for up to 2500 min. Hence, a vertical concentration gradient was present. At t = 0, a molar fraction of 40 mol% 1-butanol was observed in the bottom voxel, while in the top voxel, a fraction of 50 mol% was observed, which is in good agreement with the Raman data. Hence, the total concentration gradient of 1-butanol could be roughly estimated as 10 mol%. As the droplet height at t = 0 was approximately 600 µm, the concentration gradient equalled 0.017 mol%/µm. This gradient was generated on a timescale of several minutes, as no measurements could be carried out due to a transfer dead time between droplet deposition outside the NMR spectrometer, inserting the sample, preparing the spectrometer and the start of the first measurement.

At approximately 500 min, at which time the 1-butanol fraction in the top voxel decayed to zero, a molar fraction of 20 mol% 1-butanol was observed in the bottom voxel. For an assumed droplet height of 500 µm, this equalled a gradient of 0.04 mol%/µm. For both time points, the observed concentration gradients were of the same order of magnitude as those observed with Raman; albeit the evaporation rates were different due to the different measurement environments.

For the bottom voxel, the 1-butanol fraction rose at 1250 min. We ascribe this observation to a change in the droplet shape at this point in time, which induced a change in the concentration field within the droplet.

During evaporation, the liquid-gas interface migrated from top to bottom so that the interface after a certain time was first in the upper voxel, later in the middle and finally in the lowest voxel. This had two consequences. First, the amount of material in the voxel decreased as soon as the interface passed through the voxel. This caused the signal intensity to decrease and the signal-to-noise ratio to decrease. On the other hand, as described above, magnetic field inhomogeneities occurred at the interface, which had negative effects on the signal shape. Both a poor signal-to-noise ratio and wide line shapes can have a negative effect on the determination of the composition from the relative signal integrals. Therefore, as soon as the interface reached the top voxel, this voxel was shifted downwards so that it overlapped with the middle voxel. From the moment the interface was in this shifted voxel, only the middle and bottom voxels were considered. The error (coloured band in figure 4) was estimated from a PRESS spectrum of pure 1-hexanol (see figure S13). These errors of course propagated into the calculated concentration gradients, so the values given above are estimates.

Comparing the evaporation times of the same drops with the three methods, it is evident that the evaporation times were very different, although the same mixtures and the same liquid volumes were used for the droplets. The droplets that were investigated by Raman spectroscopy evaporated in the shortest time, and those investigated with NMR spectroscopy took the longest time. For the contact angle measurements, the droplets showed an intermediate evaporation time. This discrepancy was caused by the different surroundings and ambient conditions. The spatially resolved NMR spectroscopy measurements were taken in tubes, resembling a nearly closed system, while the non-spatially resolved NMR spectroscopy measurements were taken with air flow through the probe head. For contact angle goniometry, an ambient chamber with tuneable humidity conditions was used. The drops investigated by Raman spectroscopy were not surrounded by any chamber, and the humidity was not held constant in the laboratory.

**Discussion**
With both complementary imaging spectroscopy methods, Raman SIP and spatially resolved NMR, the evolution of the concentration gradients in an evaporating binary droplet was characterized for the first time. The challenge of discriminating two chemically very similar components, here two alcohols, in vibration spectroscopy was solved by heavy isotope labelling (stable isotope probing). In contrast to previous methods, no additional markers, e.g., fluorophores, which may be surface active themselves, are needed. Using $^2$H isotopologues of one component influences evaporation properties to a neglectable extent, as long as strong intermolecular interactions such as hydrogen bonds are retained by using proper isotopologues such as 1-butanol-$d_9$. For the NMR experiments employed here, isotope labelling is not necessary; hence, this method can be considered a "marker-free" method. Raman SIP delivers concentration information with a high temporal and spatial resolution. Compared to Raman spectroscopy, NMR yields a lower spatial and temporal resolution, but the high atom specificity and very large variety of available NMR experiments, such as diffusion ordered spectroscopy, MR imaging with different contrasts, relaxation measurements and multidimensional experiments, make NMR a very promising method for examining drying processes, e.g., molecular diffusion changes during evaporation of droplets due to changes in viscosity. Additionally, with NMR, non-transparent liquids can be examined.

Overall, Raman and NMR can be considered a promising pair of complementary methods for the examination of evaporating droplets, as they are based on two different physical effects and show different spatiotemporal resolutions; both can be considered noninvasive and (nearly) marker free.

The concentration measurements carried out here were performed for a binary droplet of 1-butanol/1-hexanol on a hydrophobic PFDTS surface. The contact angle measurements show evaporation behaviour with a mixed mode between the CCA and CCR modes. Raman spectroscopy measurements indicate that evaporation rates are enhanced at the three-phase contact line. Both methods reveal that 1-butanol evaporates faster from the droplet, generating concentration gradients of up to 0.04 mol%/µm inside the sessile droplet. Both Raman and NMR measurements show that these gradients develop on a fast time scale, as for both methods, a gradient is observed in the first measurement after droplet deposition. These concentration gradients give rise to a surface tension gradient at the surface of the droplet, which induces Marangoni flow. The evolution of the concentration gradient can be obtained by Raman SIP imaging.

Our approach to measuring concentration profiles with both methods opens the door to characterizing the evaporation process of miscible binary mixtures with different densities, surface tensions, vapor pressures and boiling points. Thus, Raman SIP imaging and spatially resolved NMR are powerful tools for investigating mass transport processes in binary liquid mixtures not only in droplets but also in further applications where concentration profiles need to be characterized in microfluidics devices or polymer systems.

**Materials and Methods**

**Materials**

1-Hexanol (CAS: 111-27-3, reagent grade by Sigma-Aldrich, St. Louis, Missouri, United States), 1-butanol (CAS: 71-36-3, 99 % by Alfa Aesar, Haverhill, Massachusetts, United States), and 1-butan-$d_9$-ol (CAS: 25493-17-8, 99 % grade by Sigma-Aldrich, St. Louis, Missouri, United States) were used as received.

For all measurement techniques, the substances were mixed at 50:50 mol%. The droplet volume for all measurements was set to 4.2 µL, and the droplets were placed on the substrates using 10 µL Hamilton syringes (Hamilton Company, Nevada, USA).

Silanized glass slides were used as substrates. For that purpose, 1H,1H,2H,2H-Perfluorodecyltrichlorosilane (PFDTS, CAS: 78560-44-8, abcr GmbH, Karlsruhe, Germany) was coated onto glass in a low-pressure chemical vapor deposition process that was designed to match the conditions of Mayer et al. (32) The protocol of all steps needed for silanization can be found in the SI of Hartmann and Hardt. (33)

**Methods**

**Raman Spectroscopy**

A confocal Raman microscope (alpha 300R, Witec GmbH, Ulm, Germany) with a green laser (532 nm, Nd:YAG) at 1 mW laser power and a Nikon 10x/0.25 objective was used for the Raman spectroscopic concentration measurements. For the concentration measurements, depth image scans were performed with a size of 90x1000 µm$^2$ (xz-plane) in an upwards direction. The integration time was 0.3 s per pixel with 9x100 pixels in each scan. These scans were repeated continuously every 8.5 min. A sketch of the experimental setup for the Raman experiments is given in figure S7.

The background was subtracted, and cosmic rays were removed from the spectra by the software WITec Project 5 (WITec GmbH, Ulm, Germany) before cross-sections were taken in the z-direction to evaluate the concentration profiles. To calculate the concentration, the peak ratio between the Raman bands at 2900 cm$^{-1}$ (protonated substance) and 2150 cm$^{-1}$ (deuterated substance) was used, which was then correlated to the calibration curve of the respective mixture. These calibration curves were measured with point scans in the mixture with integration times of 3 x 5 s at 5 mW laser power for every defined mixture. All depth measurements within the binary mixture droplets were performed beside the centre of the drop to find a possible concentration gradient. In the centre of the droplet, the concentration should remain constant over the height. (10, 12)The Raman experiments were performed under laboratory conditions (22.0 to 23.5 °C and 40 to 60 % RH). Due to the lens effect of the droplet, the droplet height was recalculated as explained in the supporting information (figure S8 and S9).

**Nuclear Magnetic Resonance Spectroscopy**

NMR spectra (Figure 2, SI Figure S12) and MR images (Figure 3) were acquired with a Bruker Avance III HD spectrometer (400 MHz proton resonance frequency, 9.4 T) equipped with a narrow bore micro 5 probe with x, y, z gradients and three GREAT 60 (A) gradient amplifiers. (23) All images and NMR spectra were acquired using an insert with a 10 mm coil tuned to the resonance frequency of $^1$H at room temperature and without spinning of the sample. MRI and NMR data were acquired and processed using Paravision 6 and Topspin 3.1PV, respectively. For all experiments, standard sequences for FLASH (34, 35), RARE (36), and PRESS (37, 38) from the Bruker pulse sequence library were used. The acquisition parameters for the MR images and NMR spectra are given in the figure captions.

All samples were prepared in 10 mm NMR tubes (Wilmad Labglas HT WG-4000. (21) Acrylate inserts were 3D printed on a Pico2$^{HD}$ 27 (Asiga, Erfurt, Germany) stereolithographic 3D printer using PlasCLEAR (Asiga, Erfurt, Germany) or Clear Resin BV-007 (MIICRAFT) photoresin. Hydrophobic surfaces were used as described above. These 7 mm round microscopy glasses (Plano) were glued to the 3D printed inserts with cyanoacrylate superglue (Pattex, Henkel).

Additionally, NMR spectra of the entire droplet (Figure S10) were recorded in a 200 MHz (4.7 T) magnet using a custom-built probe and applying a Hahn Echo pulse sequence. With the use of a weaker magnetic field, it is assumed that magnetic field distortions due to different magnetic susceptibilities of the substrate, liquid and air have a smaller impact on the resulting spectra. The in-house-built probe had a sample holder for a substrate and a solenoid RF coil similar to that presented in previous work. (22) Droplets with sizes up to a couple of microlitres could be investigated, and a continuous air flow of approximately 5 litres/min was applied.

**Acknowledgments**

This study was funded by the Deutsche Forschungsgemeinschaft (DFG, German Research Foundation) – Project ID 265191195 – SFB 1194, "Interaction between Transport and Wetting Processes", Projects A02, A06, A07 and A08. We thank T. Gambaryan-Roisman and M. Heinz (Project A04 within CRC 1194) for sharing their unpublished Droplet EvolutioN: Impact, Imbibition, Spreading and Evaporation (DENIISE) MATLAB algorithm.

**Figures and Tables**

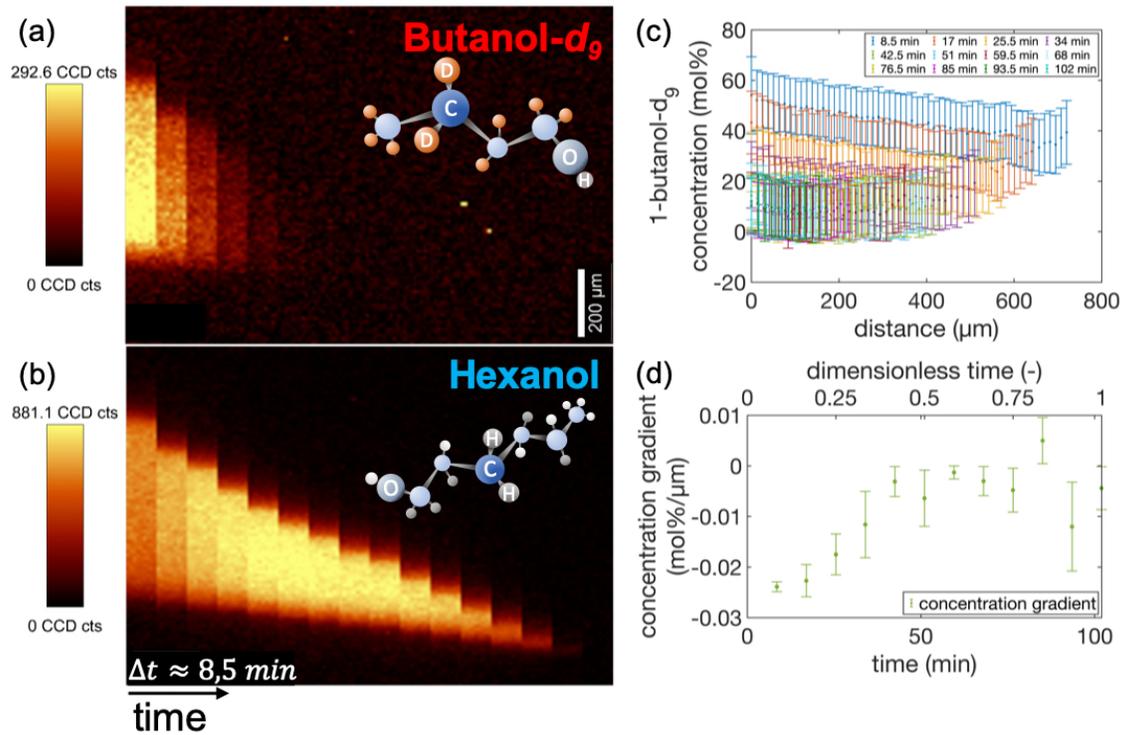

**Figure 1.** Temporal sequences of depth scan images filtered for the (a) deuterated substance and (b) protonated substance. (c) Calculated concentration profiles for each scan, and (d) 1-butanol-$d_9$ concentration gradient over experimental time at a position slightly off-centre of the droplet.

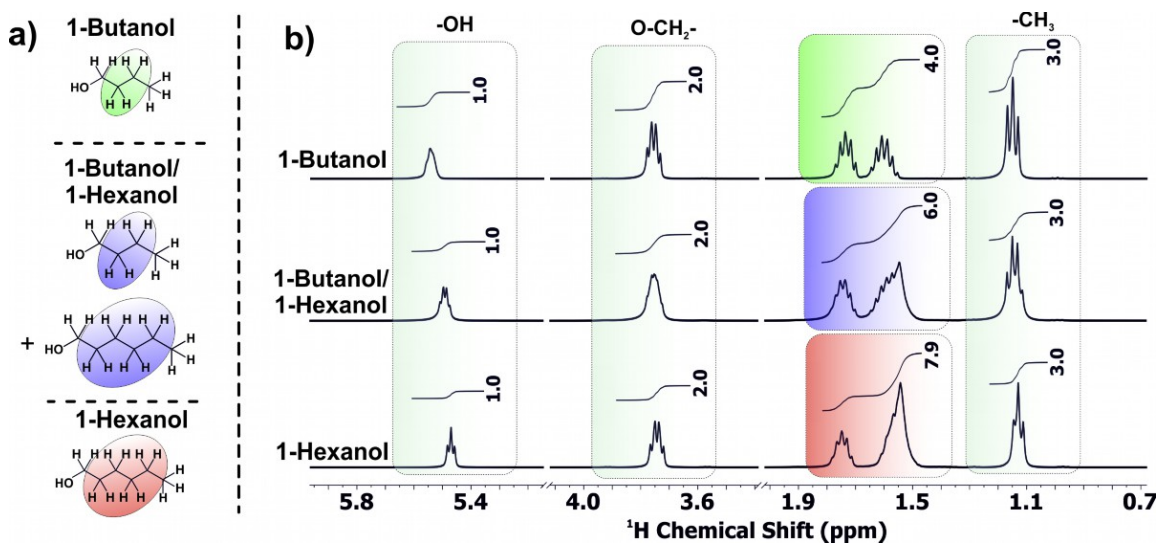

**Figure 2.** (a) Chemical structures of 1-butanol and 1-hexanol. $CH_2$ resonances are highlighted. b) $^1$H NMR spectra (400 MHz, 300 K) of pure 1-butanol, a 50 mol% mixture of 1-butanol and 1-hexanol, and pure 1-hexanol measured as isotropic bulk samples in 5 mm NMR tubes. The NMR spectra were acquired with a single scan and 64 k data points over a sweep width of 20 ppm within an acquisition time of 4.089 s. For all three spectra, the integrals of the hydroxy groups (-OH), methylene groups near the oxygen (O-$CH_2$) and methyl groups (-$CH_3$) show the same value (grey boxes). For pure 1-butanol, for the remaining methylene groups (-$CH_2$, green box), a signal integral of four is obtained. For pure 1-hexanol, the remaining $CH_2$ groups have a signal integral of eight (red box). Hence, for a 50:50 mol% mixture of these two alcohols, a signal integral of six (blue box) is observed.

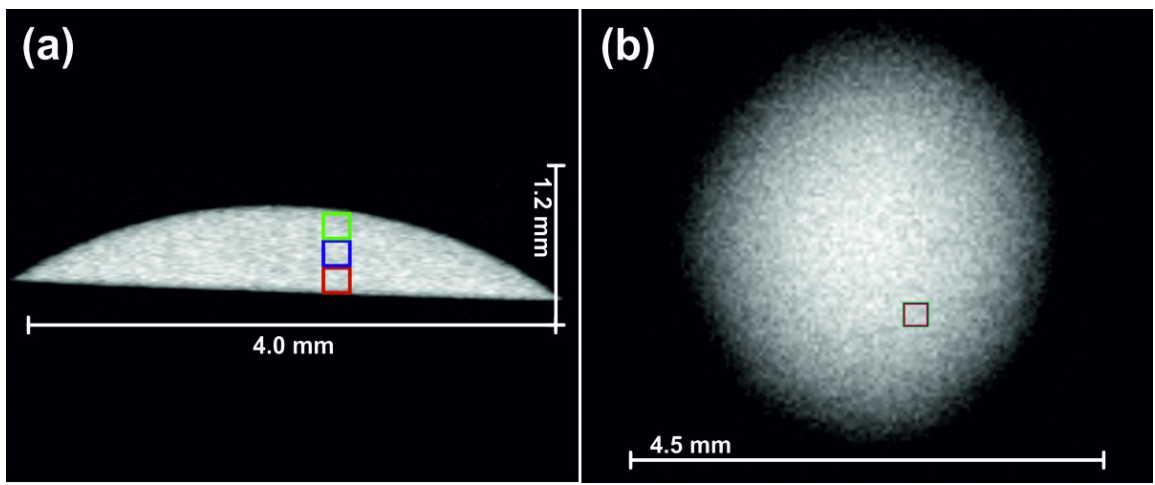

**Figure 3.** a) Sagittal RARE VTR (Rapid Acquisition with Relaxation Enhancement with variable repetition time TR) image and b) axial RARE (Rapid Acquisition with Relaxation Enhancement) image (36) of a fresh 1-butanol/1-hexanol droplet on a PFDTS surface. Both images are acquired at 400 MHz proton resonance frequency (9.4 T). The positions of the voxels used for PRESS are highlighted with coloured boxes. Both images were acquired with a 128 x 128 matrix with a field of view of 4.5 x 4.5 mm and a slice thickness of 1 mm. For the sagittal image, a single scan with an echo time (TE) of 20 ms, a repetition time (TR) of 1000 ms and a RARE factor of 8 was used. For the axial image, two scans with TE = 20.83 ms and TR = 2000 ms and a RARE factor of 8 were accumulated.

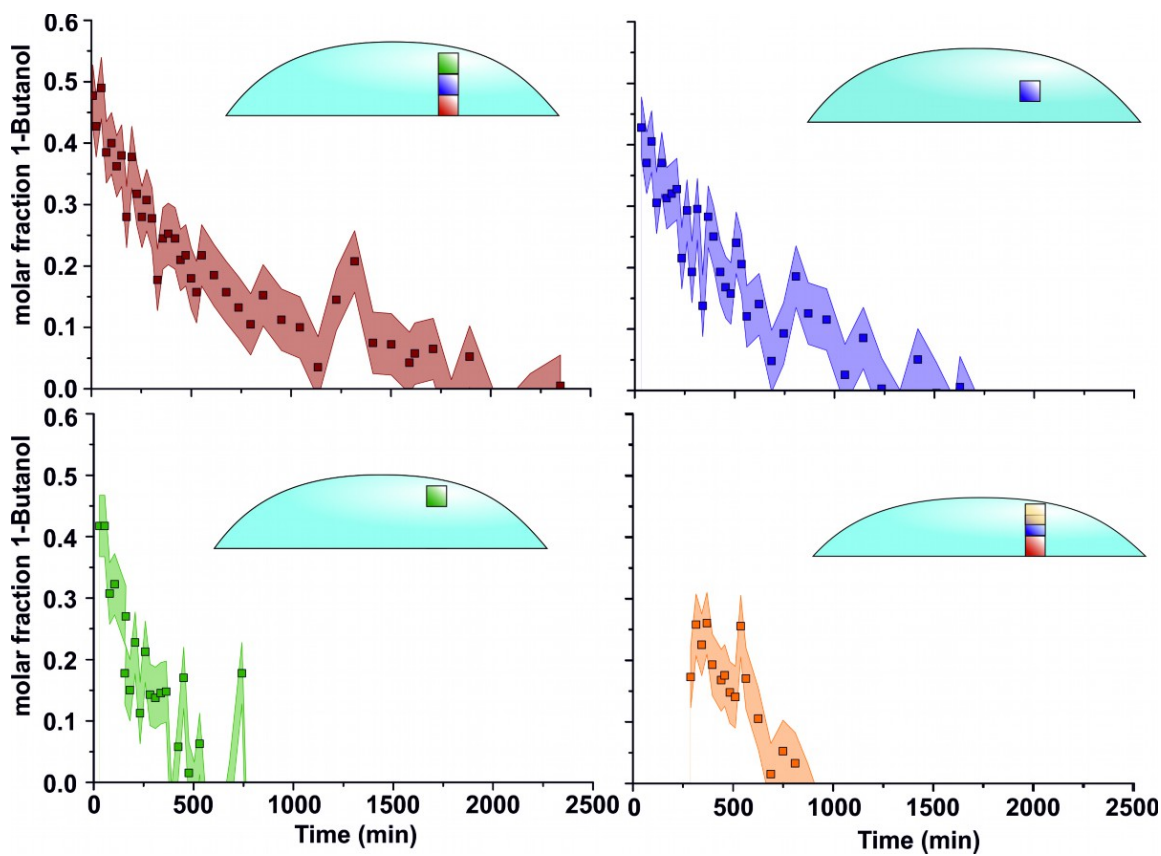

**Figure 4.** Fraction of 1-butanol in four different voxels (0.2 x 0.2 x 0.2 mm) during evaporation over time t at 298.5 K. The voxel positions are indicated within the drawing. The PRESS NMR spectra were acquired with 64 scans, a repetition time of 4000 ms and an echo time TE of 20 ms. For each FID, 8000 data points were acquired within an acquisition time of 998.4 ms. The orange voxel is shifted downwards compared to the green voxel to avoid a moving liquid/gas interphase within the voxel. The 1-butanol fraction is calculated by comparing the signal integrals of $CH_2$ and $CH_3$ resonances.